\documentclass{PoS_arXiv}
\usepackage{cite, amsmath}

\title{Higher-Order Contributions in Higgs Sectors of Supersymmetric Models}

\ShortTitle{Higher-Order Contributions in Higgs Sectors of SUSY Models}

\author{Thomas Hahn\\
        Max-Planck-Institut f\"ur Physik, 
        F\"ohringer Ring 6,
        80805 M\"unchen, Germany\\
        E-mail: \email{hahn@feynarts.de}}

\author{Sven Heinemeyer\\
        Instituto de Fisica de Cantabria (CSIC-UC), 
        Santander, Spain\\
        E-mail: \email{Sven.Heinemeyer@cern.ch}}

\author{Wolfgang Hollik\\
        Max-Planck-Institut f\"ur Physik, 
        F\"ohringer Ring 6,
        80805 M\"unchen, Germany\\
        E-mail: \email{hollik@mppmu.mpg.de}}

\author{\speaker{Heidi Rzehak}\\
        Albert-Ludwigs-Universit\"at Freiburg, Physikalisches Institut,
        Hermann-Herder-Str. 3, 79104 Freiburg, Germany\\
        E-mail: \email{heidi.rzehak@\,\!physik.uni-freiburg.de}}

\author{Georg Weiglein\\
        DESY, Notkestra\ss e 85, 22607 Hamburg, Germany\\
        E-mail: \email{Georg.Weiglein@desy.de}}
\abstract{In 2012, the discovery of a particle compatible with a Higgs boson
  of a mass 
of roughly 125~GeV was announced. This great success is now being followed by
the identification of the nature of this particle and the particle's
properties are being 
measured. One of these properties is the Higgs boson mass which is already
known very precisely with an experimental uncertainty of below 1~GeV. In some
extensions of the Standard Model, like in supersymmetric extensions, the Higgs
boson mass can be predicted and  hence, the measured mass constrains the
parameters of the model. For a full exploitation of this constraint, a precise
theoretical prediction is needed. The presented combination of the results
obtained by the Feynman diagrammatic approach and the renormalization group
equation approach 
improves the known Higgs mass prediction for larger mass scales of the
superpartner particles. 
}

\FullConference{ Loops and Legs in Quantum Field Theory - LL 2014,\\
		 27 April - 2 May 2014 \\
		 Weimar, Germany }

\begin{document}

\section{Introduction}
The biggest success of the first run of the LHC is the discovery of a
particle in the Higgs search channels \cite{ATLASdiscovery, CMSdiscovery}. The properties of this particle are
being measured and conform so far with the Higgs boson of the Standard
Model (SM). In the SM, only one Higgs boson exists and its mass is a
free parameter. In extensions of the SM, however, several Higgs
bosons might exit with one of them being the discovered one and their masses
can  depend on other parameters of the 
theory. Supersymmetric models like the Minimal Supersymmetric Model (MSSM) or
the Next-to MSSM fall into this category. In the MSSM, the mass of the
lightest Higgs boson can be predicted from other parameters in the
theory.  In this case, the mass measurement of the newly discovered boson
 can be exploited to constrain the parameters of the underlying theory. 

\section{The Higgs Sector in the MSSM}

The Higgs potential $V_{\mathrm{Higgs}}$ in the MSSM is given by
\begin{align}\nonumber
V_{\mathrm{Higgs}}
&=   \frac{g^2 + {g'}^2}{8} (H_d^{+} H_d
  - H_u^{+} H_u)^2 + \frac{{g}^2}{2} |H_d^{+} H_u|^2
\\[0.3cm]& \quad\ \nonumber
+  | \mu|^2 (H_d^{+} H_d + H_u^{+} H_u) 
\\[0.3cm]& \quad\  
 + (m_1^2 H_d^{+} H_d + m_2^2
H_u^{+} H_u)
%\\[0.3cm]& \quad\ \nonumber
 + (\epsilon_{ij} |m_{12}^2| e^{ i  \varphi_{m_{12}^2}}
H_d^i H_u^j + h.c.)~.
\label{eq:HiggsPot}
\end{align}
While in the SM the parameters of the Higgs potential do not appear in the
other sectors of the model the MSSM Higgs potential contains terms depending 
on the gauge couplings $g$ and $g'$ (as can be seen in the first line of
Eq.~\eqref{eq:HiggsPot}) due to the underlying
supersymmetry. Furthermore, a term proportional to the absolute value squared
of the coupling  
$\mu$ governing the mixing of the Higgs superfields exists. Finally, the soft
breaking part of the Higgs potential depends on the soft breaking parameters 
$m_1^2$, $m_2^2$ and $m_{12}^2$. The latter of these parameters can be complex
in general, however, applying a Peccei-Quinn transformation this phase can be
rotated away without changing the physical content of the model~\cite{CPphases}.
 A second
phase can appear as a phase difference of the two vacuum expectation
values. Exploiting the minimum condition for the potential leads, however, to a
vanishing phase. Hence, there are no physical phases at the tree-level in the
Higgs sector of the MSSM and hence no CP-violation.

In the MSSM, five physical Higgs bosons with three neutral and two
charged ones exist where two of the neutral ones are CP-even and one is CP-odd 
in
the absence of CP-violation. The masses of these Higgs bosons are not all
independent. To be more precise, there are two independent parameters in the
Higgs sector which are often chosen as the mass of the CP-odd Higgs boson or
the charged Higgs boson mass in a CP-conserving or a CP-violating scenario,
respectively, and the ratio of the Higgs vacuum expectation values $\tan \beta
= v_u/v_d $. Hence, three of the four masses can be predicted. The mass of
the lightest Higgs boson has an additional theoretical upper limit which, at
Born level, is at the Z~boson mass $M_h \leq M_Z$ but is shifted to higher
values via quantum corrections so that $M_h \lesssim 135$~GeV for masses of
 the supersymmetric top quark partners of
up to a couple of TeV~\cite{theoryer}. The mass of the
discovered particle with $M_h \approx 125.5$~GeV lies within this range and
the lightest Higgs boson could describe the discovered particle (see e.g. \cite{Hscans}). Due to the
dependence of the quantum corrections on the MSSM parameters the Higgs boson
mass is an interesting precision observable---however, to completely exploit
this information the uncertainty of the theoretical prediction should at least
match the experimental accuracy. While the experimental measurement of the
Higgs mass $M_h$ is already very precise with 
\begin{align} \nonumber 
\text{ATLAS:}& \quad M_h = 125.5 \pm 0.2 (\text{stat}) + {}^{+0.5}_{- 0.6} (\text{sys})~\text{GeV~~\cite{MhATLAS}}\\\nonumber
\text{CMS:}& \quad M_h = 125.7 \pm 0.3 (\text{stat}) \pm 0.3 (\text{sys})~\text{GeV~~\cite{MhCMS}}
\end{align}
and hence an experimental error of
$\Delta M_{h}^\text{exp.} < 1$ GeV the
estimated theory uncertainty is approximately 
$\Delta M_{h}^\text{theory} \approx 3$~GeV~\cite{theoryer}. 
An improvement on the theory uncertainty is therefore mandatory.
 
Special questions that have been discussed in the context of Higgs mass 
constraints are how light
the top quark partners, the top squarks, can be, and how large the mixing
between the partners of the left-handed and the right-handed top
quark must be in order to yield a Higgs mass value matching the measured one.

Not only for constraining the parameter space, a precise prediction of the 
Higgs boson mass is needed but also as an
input for the consistent calculation of Higgs production cross sections and
partial decay widths within the MSSM.

%The predicted MSSM Higgs boson mass has additionally drawn attention  with
%respect to the discussion of the amount of fine-tuning which 
%is needed to shift the Higgs boson mass from the Born-mass value to the
%measured one. Questions that have been discussed in this context are how light
%the top quark partners, the top squarks, can be, and how large the mixing
%between the partners of the left-handed top quark and the right-handed top
%quark must be in order to yield a Higgs mass value matching the measured one.

\section{Calculation of the Higgs masses in the MSSM}

Two main methods have been applied for the calculation of the Higgs boson
masses in the MSSM: the Feynman diagrammatic (FD) approach\footnote{The
  effective 
  potential approach and the FD approach in the
  approximation of vanishing external momenta lead to the same
  result.}~\cite{1loopFD, 1loopFDFH, 2loopFDFH, 2loopFD, 2loopFD_2, H3m}  
and the renormalization group equation (RGE)
approach~\cite{1loopRGE,RGE2l_1,RGE2l_2, RGE2l_3, RGE2l_4, RGE3l_1, RGE3l_2}.

\subsection{The Feynman-diagrammatic approach}
 
In the FD approach, Feynman diagrams that contribute to the
renormalized Higgs 
self energies $\hat{\Sigma}$ are calculated and the renormalized two-point
vertex function 
$\hat{\Gamma}$
\begin{align}
- \text{i} \hat{\Gamma} (p^2) = p^2 - {\bf M} (p^2)
\end{align}
has to be determined where $M(p^2)$ is the loop-corrected Higgs-mass matrix
\begin{align}
 {\bf M} (p^2) =   \begin{pmatrix}  
M_{h_{\text{Born}}}^2-\Hat{\Sigma}_{hh}(p^2)
 &-\Hat{\Sigma}_{Hh}(p^2) 
  &-\Hat{\Sigma}_{hA}(p^2)\\
   -\Hat{\Sigma}_{Hh}(p^2)
& M_{H_{\text{Born}}}^2 -\Hat{\Sigma}_{HH}(p^2)
& -\Hat{\Sigma}_{HA}(p^2)\\
-\Hat{\Sigma}_{hA}(p^2) & -\Hat{\Sigma}_{HA}(p^2) &
M_{A_{\text{Born}}}^2-\Hat{\Sigma}_{AA}(p^2)
\end{pmatrix} \label{eq:massmatrix}
\end{align}
with the Born masses squared $M_{h_{\text{Born}}}^2$, $M_{H_{\text{Born}}}^2$
and $M_{A_{\text{Born}}}^2$ for the light CP-even, the heavy CP-even and the
CP-odd Higgs boson, respectively, and the renormalized Higgs self energies
$\hat{\Sigma}_{\phi \chi}$ with $\phi, \chi = h, H, A$. 

In the case of real parameters and hence CP-conservation, the mixing between
CP-odd and CP-even Higgs bosons vanishes,
$\Hat{\Sigma}_{hA}(p^2) = \Hat{\Sigma}_{HA}(p^2) =0$.

The calculations of the zeroes of the determinant of the two-point vertex function  
$\det(\Gamma) = 0$ yields the values for the loop-corrected Higgs boson
masses.

\subsection{The renormalization group equation approach}

In the simplest version of the RGE approach it is
assumed that all the supersymmetric partner particles are heavy with masses of
the order of the mass scale $M_S$. At energies larger than the mass scale
$M_S$ the full MSSM theory is active while below the SM as an
effective theory is a good
description.  The effective theory is matched to the MSSM
via the requirement that the quartic Higgs coupling in the MSSM
$\lambda^{\text{MSSM}}$ and in the SM $\lambda^{\text{SM}}$ adopt
the same value at the scale $M_S$,
\begin{align}
 \lambda^{\text{MSSM}}(M_S) = \lambda^{\text{SM}} (M_S)
\end{align}
Starting from this value of $\lambda^{\text{SM}}$, the SM RGEs are used to
evolve $\lambda^{\text{SM}}$ to lower values. The Higgs 
mass squared at the scale of the top quark mass $m_t$ is then given as 
\begin{align}\label{eq:Hmass2}
M_h^2(m_t^2) = 2 \lambda^{\text{SM}}(m_t) v^2
\end{align} 
with  the vacuum expectation value $v \approx 174$~GeV. In general, in order to
obtain the pole mass a conversion 
between the 
$\overline{\text{MS}}$ mass of Eq.~\eqref{eq:Hmass2}  and the on-shell mass is
necessary.

\subsection{Comparison of both approaches}

Both of the approaches have their advantages:
\begin{itemize}
\item The FD approach, on the one hand, takes all logarithmic and
non-logarith\-mic 
terms into account at a certain order of perturbation theory. This is
especially important for lower mass scales. Additionally, different mass
scales are automatically implemented which, however, can lead to large
logarithmic contributions if the splitting between the mass scales is 
large. Furthermore, the calculated self-energies are not only needed for the 
mass calculation but also for the evaluation of the mixing of the Higgs bosons at 
loop-level.
\item
The RGE approach, on the other hand, has the
advantage of resumming potentially large logarithmic terms to all orders 
which is
important if large mass scales are playing a role.
\end{itemize}
In order to profit from both of these advantages we perform a combination of
both approaches~\cite{Mhresum}.

\section{Combination of both approaches}

In the following we restrict ourselves to the CP-conserving case and assume
that the parameters are real. 

The result of the FD part is adopted from the program
{\tt FeynHiggs2.9.5}~\cite{ theoryer, 1loopFDFH, 2loopFDFH, FeynHiggs}. For
the calculation of the RGE part  
the SM renormalization group equations 
at the two-loop level with vanishing electroweak gauge
couplings~\cite{RGE2l_1} have been applied. The matching is performed at the
scale   
$M_S = \sqrt{m_{\tilde{t}_1}m_{\tilde{t}_2}}$, the geometric mean of the two
top squark masses $m_{\tilde{t}_1}$, $m_{\tilde{t}_2}$ and the quartic
coupling at this scale is given
by~\cite{RGE2l_2, RGE2l_4, Matching} 
\begin{align} 
\lambda(M_S) = \frac{3 y_t^4}{8 \pi^2} \frac{X_t^2}{M_S^2}\left[1 - \frac{X_t^2}
{M_S^2}\right] \label{eq:lambdamatched}
\end{align}
where $y_t$ is the Yukawa coupling and $X_t = A_t - \mu \cot \beta$ the top
squark mixing parameter with the soft-breaking trilinear coupling $A_t$. It should be noted that also in Eq.~\eqref{eq:lambdamatched} vanishing electroweak gauge couplings have been assumed which leads to a vanishing tree-level contribution. Hence, applying Eq.~\eqref{eq:Hmass2} results in a pure higher-order correction to the Higgs mass.
Following this procedure yields a Higgs-mass prediction with the leading and the next-to leading logarithmic contributions $\propto L^n$, $\propto L^{n-1}$ 
with $L = \log(M_S/m_t)$ at n-loop order resummed to all orders.

In order to combine the FD and the RGE approach great care
has to be taken to avoid double counting of logarithms. Therefore, the
logarithmic part has to be subtracted from the FD result. 
The FD calculation is performed with a top quark mass in the
$\overline{\text{MS}}$~scheme and 
top squark masses and mixings in an on-shell scheme
while the RGE 
method leads to a result in the $\overline{\text{MS}}$ scheme throughout. To
avoid 
double counting, the logarithmic part which is subtracted has to be in the
on-shell scheme. Otherwise logarithms stemming from the translation between the
schemes will reintroduce a double counting. In the case of the mass parameter
$M_S$ no logarithms appear in the conversion from the $\overline{\text{MS}}$
 to the on-shell scheme while the conversion of the mixing parameter involves a
 logarithm and is given by
\begin{align}
X_t^{\overline{\text{MS}}} =  X_t^\text{OS}\left[1 +
    \ln\frac{M_S^2}{m_t^2}\left(\frac{\alpha_s}{\pi} - \frac{3 \alpha_t}{16
      \pi}\right)\right]~. 
\end{align}
The Higgs-mass correction is then obtained as 
\begin{align}
\Delta M_h^2 = (\Delta M_h^2)^{\text{FD}}(X_t^{\text{OS}}) - (\Delta
M_h^2)^{\text{1l,2l-log}}( X_t^{\text{OS}}) + (\Delta
M_h^2)^{\text{RGE}}( X_t^{\overline{\text{MS}}})
\end{align}
where $(\Delta M_h^2)^{\text{FD}}$, $(\Delta
M_h^2)^{\text{1l,2l-log}}$ and $(\Delta
M_h^2)^{\text{RGE}}$ are the contribution of the
FD approach, the logarithmic part of the one- and the two-loop
order $\mathcal O(\alpha_t \alpha_S)$ and $\mathcal O (\alpha_t^2)$ with parameter $X_t$ on-shell and the
part obtained from the RGE approach according to Eq.~\eqref{eq:Hmass2},
 respectively. In the case of large CP-odd
Higgs-boson masses with $M_A \gg M_Z$, the self energy of the CP-even 
Higgs-interaction
eigenstate $\phi_u$ which couples to up-type quarks is approximately
\begin{align}
\hat\Sigma_{\phi_u\phi_u} \approx (\sin \beta)^{-2} \Delta M_h^2.
\end{align}
Using this approximation the corrections can be incorporated into the
loop-corrected Higgs-mass matrix~Eq.\eqref{eq:massmatrix} and hence be taken
into account in the loop-corrected Higgs mixing, see e.g.\ \cite{1loopFDFH}.

\section{Results}

\begin{figure}[b]
\begin{tabular}{ccc}
%\begin{tabular}{cccc}
% \hspace*{1cm} &
\includegraphics[width=0.47\linewidth]{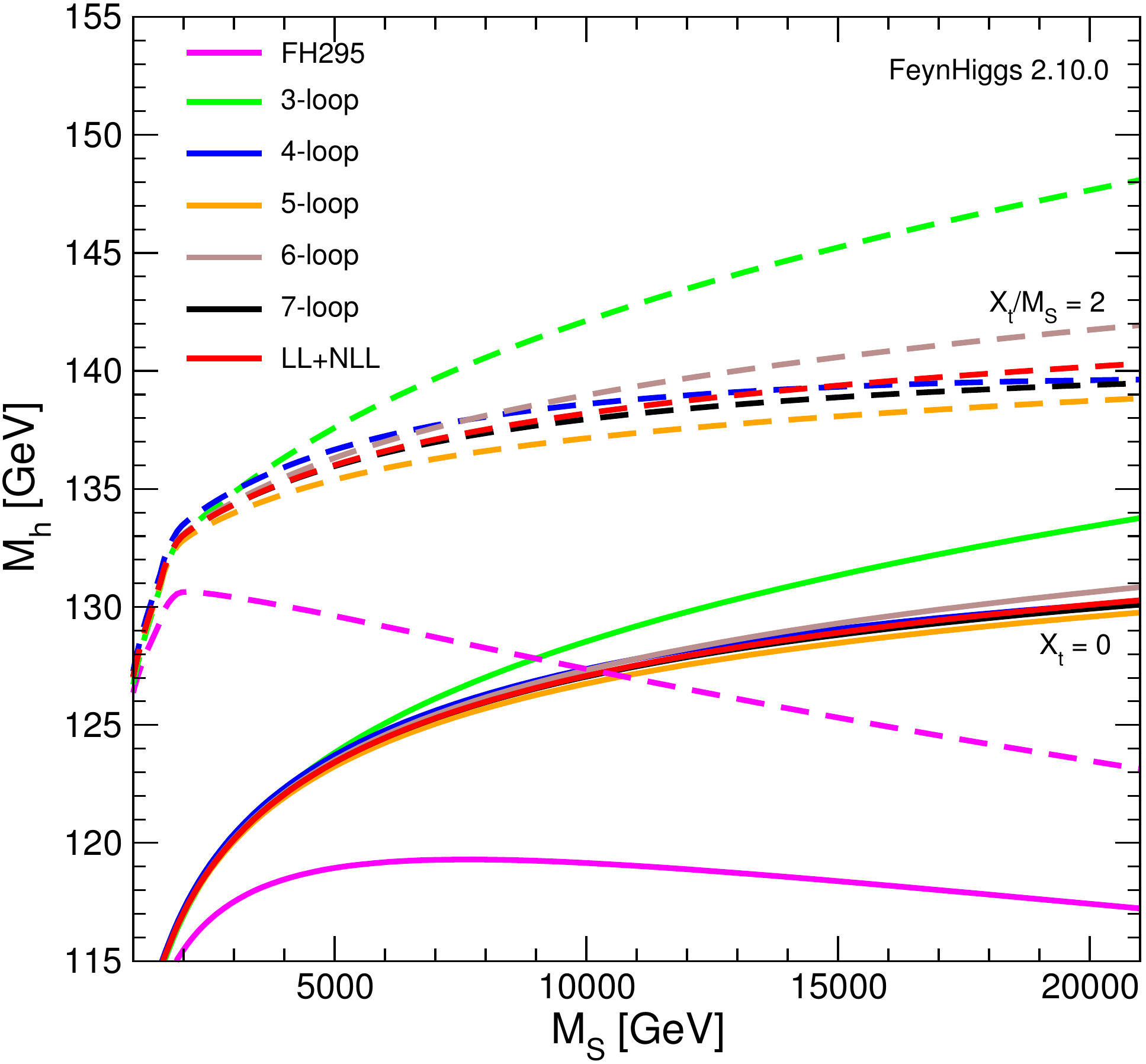}
&%\hspace*{0.1ex}
&
%& \hspace*{1cm} &
\includegraphics[width=0.47\linewidth]{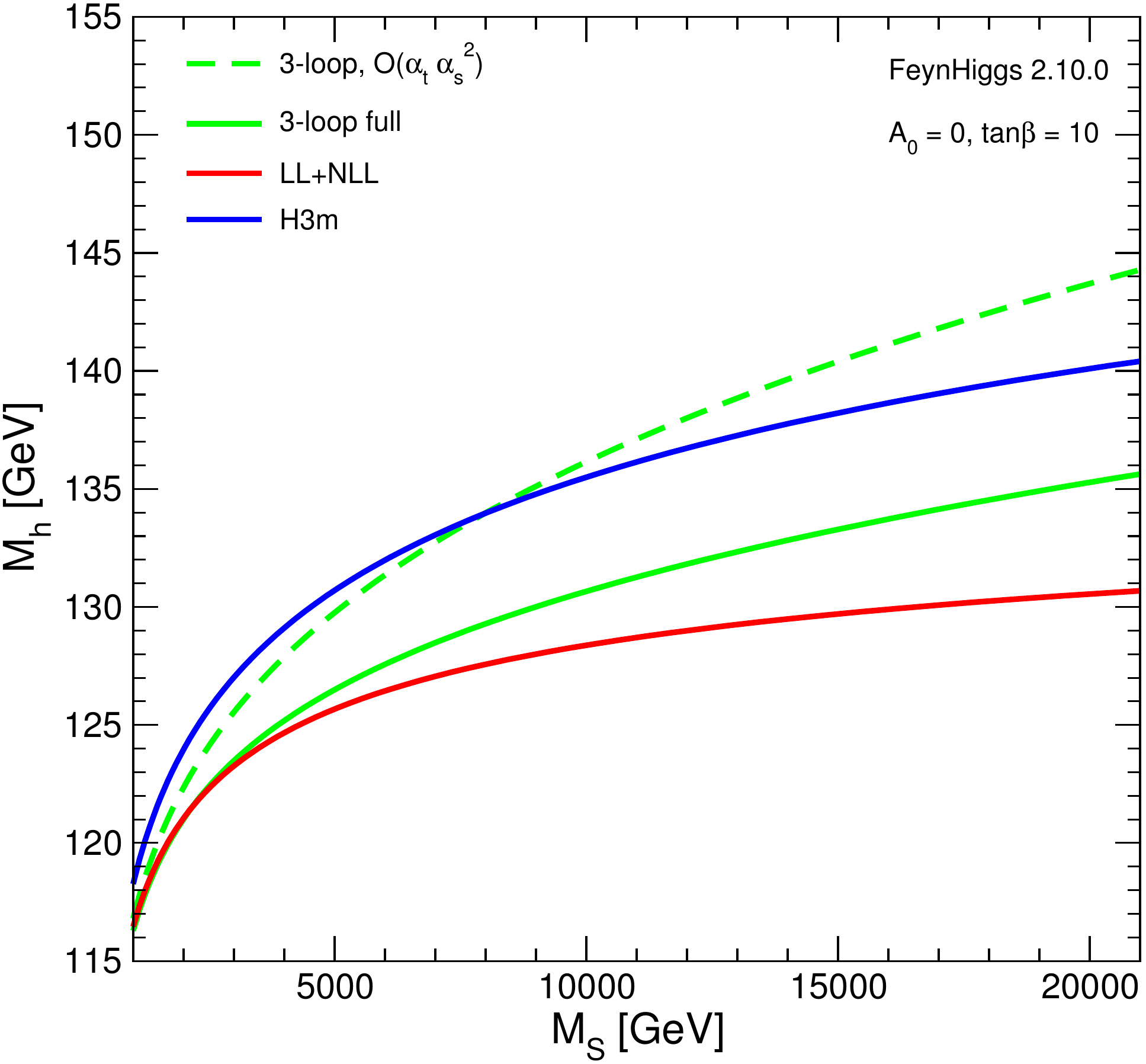}
\end{tabular}
\caption{The mass of the lightest MSSM Higgs boson $M_h$ in dependence on the mass scale $M_S$.\newline
Left: The ``old'' {\tt FeynHiggs} result with no additional logarithmic contributions of 3- or higher loop order (pink), including analytically determined leading and next-to leading logarithmic contributions of 3- to 7-loop order (green, blue, yellow, brown, black) and including the numerically obtained resummed leading and next-to leading contributions (red) are shown for no mixing/maximal mixing (solid/dashed).\newline
Right: The {\tt H3m} result is shown in blue, the {\tt FeynHiggs} result including leading and next-to leading logarithmic contributions of 3-loop order of $\mathcal O(\alpha_t \alpha_s^2)$ and additionally of $\mathcal O(\alpha_t^2 \alpha_s)$ and $\mathcal O(\alpha_t^3)$ is presented in green dashed and solid, respectively. The {\tt FeynHiggs} result including the resummed logarithmic contributions is shown in red.\newline
Parameters are chosen as described in the text.
  \label{plots}}
\end{figure}

In order to discuss the importance of the resummed corrections in
Fig.~\ref{plots} on the left side the {\tt FeynHiggs2.9.5} result (FH295) is
shown (pink) in 
comparison to the results including logarithmic higher-order contributions
which are implemented in {\tt FeynHiggs2.10.0}~\cite{theoryer, 1loopFDFH,
  2loopFDFH, Mhresum, FeynHiggs}.
The leading  and the next-to leading logarithmic terms up to  3- to
7-loop order  $\mathcal O(\alpha_t^n \alpha_s^m)$, $n+m = 3,\dots ,7$  have
been calculated analytically and combined  to the {\tt FeynHiggs} result
(shown in 
green, blue, yellow, brown and black). Numerically, the leading and
next-to leading logarithmic contributions have been resummed to all orders and
the combined 
result, which corresponds to the one of the new implementation of {\tt FeynHiggs2.10.0}, 
is shown in red.

The parameters have been chosen as follows: $M_A = M_2
= \mu = 1$~TeV with $M_2$ being the SU(2) gaugino soft breaking parameter, the
gluino mass $m_{\tilde g} = 1.6$~TeV and $\tan \beta =10$. For the
solid (dashed) lines 
no mixing $X_t = 0$ (maximal mixing $X_t/M_S = 2$) is assumed.
The old {\tt 
  FeynHiggs} result is reliable up to SUSY breaking mass scales of the order
of $\mathcal O(1 \text{ TeV})$ while for higher values of $M_S$ it
underestimates the Higgs mass value. Taking into account logarithms of 3-loop
order  
yields a valid result up to $\mathcal O(5 \text{ TeV})$ in the no mixing
case. For higher values of $M_S$ the 3-loop result overestimates the resummed
result. Thus, further
logarithmic contributions have to be taken into account and can amount to mass
shifts of several GeV. 

On the right side of Fig.~\ref{plots} a comparison of the {\tt FeynHiggs} result
with the known 3-loop order prediction of the Higgs mass is performed. The 
latter is obtained with the help of the program {\tt H3m} \cite{H3m}. 
In order to
perform the comparison a CMSSM scenario has been chosen with the soft-breaking
parameters $M_0 = M_{1/2} = 200 \dots 15000$~GeV and $A_0 = 0$ and
additionally $\tan \beta = 
10$ and $\mu >0$. The low-energy parameters have then been calculated with the
spectrum generator SoftSUSY~\cite{SoftSUSY}. The following results are shown:
the resummed leading and next-to leading log result (red), the result including 
only leading and next-to leading logarithmic terms up to 3-loop order (green, solid), the result including only up to leading and next-to leading logarithmic terms of order $\mathcal O(\alpha_t \alpha_s^2)$ (green, dashed)
and the result obtained with {\tt H3m} (blue). The program {\tt H3m} takes
into account corrections up to order $\mathcal O(\alpha_t \alpha_s^2)$% making
%use of self-energy contributions from {\tt FeynHiggs}
. As it
is calculated using the FD approach it includes all
terms of this order, also non-logarithmic terms.  Also, it takes into account
different SUSY mass scales. Additionally, the applied
renormalization schemes at
two-loop order  differ in {\tt H3m} and {\tt FeynHiggs}. With respect to 
these difference the $\mathcal O(\alpha_t
\alpha_s^2)$ results of both programs agree reasonably well. However, one can
also see that the terms of $\mathcal O (\alpha_t^2 \alpha_s)$ and $\mathcal
O(\alpha_t^3)$ are important too and lead to a reduction of the size
of the Higgs mass prediction.

\section{Conclusion}

The Higgs sector of the Minimal Supersymmetric Standard Model has been
discussed. One important and easily measurable observable is the mass of the
newly discovered particle which could be the lightest Higgs boson of the
MSSM. To fully exploit this constraint, a precise prediction of the mass of the
lightest Higgs boson is necessary. For lower SUSY mass scales the Feynman
diagrammatic approach leads to a precise prediction taking into account all 
terms
of a certain order, also non-logarithmic terms. For larger SUSY mass scales
the renormalization group equation approach is advantageous as potentially
large logarithms are resummed. The presented combination of both approaches
improves the known Feynman diagrammatic results for larger SUSY mass
scales. It is implemented into the program {\tt FeynHiggs} and can be switched
on by the user by setting the flag {\tt looplevel =3}.

Further refinements, e.g. allowing for different mass scales or including 
higher-order contributions in the matching conditions, are planned.

\section*{Acknowledgments}

H.R. would like to thank the organizers of ``Loops and Legs
2014'' for the invitation to present this work and for an interesting and enjoyable conference.

\end{document}